\documentclass[12pt]{article}
\usepackage{epsfig}
\usepackage{cite}
\usepackage{amsmath, amssymb, amsfonts}
\usepackage{color}
\usepackage{latexsym}
\usepackage{graphicx}
\usepackage[font=small,labelfont=bf]{caption}
\usepackage[colorlinks,bookmarks]{hyperref}
\hypersetup{pdfpagemode=UseNone, pdfstartview=FitH, linkcolor=blue,
            citecolor=red, urlcolor=blue}

\def\be{\begin{equation}}
\def\ee{\end{equation}}
\def\ba{\begin{eqnarray}}
\def\ea{\end{eqnarray}}

\def\bdm{\begin{displaymath}}
\def\edm{\end{displaymath}}
\def\la{~\mbox{\raisebox{-.6ex}{$\stackrel{<}{\sim}$}}~}
\def\ga{~\mbox{\raisebox{-.6ex}{$\stackrel{>}{\sim}$}}~}
\def\bq{\begin{quote}}
\def\eq{\end{quote}}

 at 10truept

\newcommand{\rmd}{\mathrm{d}}

\newcommand{\avg}[1]{\langle #1 \rangle}

\renewcommand{\[}{\left[}
\renewcommand{\]}{\right]}
\renewcommand{\(}{\left(}
\renewcommand{\)}{\right)}
\newcommand{\vk}{\vec{k}}

\newcommand{\eps}{\epsilon}

\newcommand{\lam}{\lambda}

\newcommand{\Om}{\Omega}

\newcommand{\Hz}{\textrm{Hz}}

\newcommand{\Mpc}{\textrm{Mpc}}

\newcommand{\Mpl}{M_{\mathrm{Pl}}}

\newcommand{\bea}{\begin{eqnarray}}
\newcommand{\eea}{\end{eqnarray}}

\newcommand{\bi}{\begin{itemize}}
\newcommand{\ei}{\end{itemize}}

\newcommand{\beq}{\begin{equation}}
\newcommand{\eeq}{\end{equation}}
\newcommand{\beqa}{\begin{eqnarray}}
\newcommand{\eeqa}{\end{eqnarray}}

\def\la{~\mbox{\raisebox{-.6ex}{$\stackrel{<}{\sim}$}}~}
\def\ga{~\mbox{\raisebox{-.6ex}{$\stackrel{>}{\sim}$}}~}

\newcommand{\calO}{\mathcal{O}}

\newcommand{\calL}{\mathcal{L}}

\def\ltap{\ \raise.3ex\hbox{$<$\kern-.75em\lower1ex\hbox{$\sim$}}\ }
\def\gtap{\ \raise.3ex\hbox{$>$\kern-.75em\lower1ex\hbox{$\sim$}}\ }
\def\gl{\ \raise.5ex\hbox{$>$}\kern-.8em\lower.5ex\hbox{$<$}\ }
\def\roughly#1{\raise.3ex\hbox{$#1$\kern-.75em\lower1ex\hbox{$\sim$}}}

\begin{document}

\thispagestyle{empty}
\begin{flushright}
January 2021 \\
DESY 21-004 \\
\end{flushright}
\vspace*{.35cm}
\begin{center}
  
{\Large \bf Double Monodromy Inflation: A Gravity Waves}
\vskip.3cm
{\Large \bf Factory for CMB-S4, LiteBIRD and LISA}

\vspace*{1cm} {\large Guido D'Amico$^{a, }$\footnote{\tt
damico.guido@gmail.com}, Nemanja Kaloper$^{b, }$\footnote{\tt
kaloper@physics.ucdavis.edu} and 
 Alexander Westphal$^{c,}$\footnote{\tt alexander.westphal@desy.de}
}\\
\vspace{.3cm} {\em $^a$Dipartimento di SMFI dell' Università di Parma and INFN}\\
\vspace{.05cm}{\em Gruppo Collegato di Parma, Italy}\\
\vspace{.3cm}
{\em $^b$QMAP, Department of Physics, University of
California, Davis, CA 95616, USA}\\
\vspace{.3cm} $^c${\em Deutsches Elektronen-Synchrotron DESY, Theory Group, D-22603 Hamburg, Germany}\\

\vspace{1.15cm} ABSTRACT
\end{center}
We consider a short rollercoaster cosmology based on two stages of monodromy inflation separated by
a stage of matter domination, generated after the early inflaton falls out of slow roll. If the first stage
is controlled by a flat potential, $V \sim \phi^p$ with $p < 1$ and lasts ${\cal N} \sim 30 - 40$ efolds, the
scalar and tensor perturbations at the largest scales will fit the CMB perfectly, and produce relic gravity waves
with $0.02 \la r \la 0.06$, which can be tested by LiteBIRD and CMB-S4 experiments. If in addition the first inflaton
is strongly coupled to a hidden sector $U(1)$, there will be an enhanced production of vector fluctuations
near the end of the first stage of inflation. These modes convert rapidly to tensors during the short epoch
of matter domination, and then get pushed to superhorizon scales by the second stage of inflation,
lasting another $20-30$ efolds. This band of gravity waves is chiral, arrives today with wavelengths 
in the range of $10^8$ km, and with amplitudes greatly enhanced compared to the long wavelength CMB modes
by vector sources. It is therefore accessible to LISA.
Thus our model presents a rare early universe theory predicting \emph{several simultaneous} 
signals testable by a broad range of gravity wave searches in the very near future.

\vfill \setcounter{page}{0} \setcounter{footnote}{0}

\vspace{1cm}
\newpage

\section{Introduction}

Inflation \cite{Guth:1980zm, Linde:1981mu,Albrecht:1982wi} arose as a paradigm to 
explain the universe {\it naturally}. It has been noted
well prior to its advent that the universe is incredibly unnatural at the largest scales \cite{Collins:1972tf}. 
Without accelerated expansion early on, generic  
initial conditions would have yielded a far less hospitable universe, which raised the questions 
of fine tuning and anthropic selection early on.
The mechanism of inflation addresses these problems by reducing the sensitivity to 
the initial conditions, metaphorically taking
the log of the measure of tuning: one needs a total of $\sim 60$ efolds to shield from bad 
influences of initial anisotropies and
inhomogeneities. As a bonus one receives a mechanism to generate structures at 
shorter scales, using local and causal dynamics
enshrined in the effective field theory (EFT) of the inflaton sector.

Thus inflation translates the problem of cosmological naturalness to the problem of 
naturalness of the inflaton EFT. The construction of 
natural EFTs is relatively straightforward in the limit of semiclassical gravity. For 
example, all one needs is a single field with a flat potential 
and derivative couplings to everything else, and a model is born \cite{Kaloper:2011jz}. 
However with full-on quantum gravity, building UV complete 
inflation models is quite challenging. One general approach which was initiated 
in the past decade or so is monodromy inflation  
\cite{Silverstein:2008sg,McAllister:2008hb,Kaloper:2008fb,Kaloper:2011jz}, 
where the natural EFTs of the inflaton can be protected from the perils of quantum 
gravity by embedding them into gauge theories 
spontaneously broken at some scale above the scale of inflation, 
but below the fundamental gravity scale. 

Even so, deploying 
monodromy models is nontrivial, with some pressure coming from both the theory side due to 
backreaction induced in setups with large field 
variations, and the observations, because the dynamics predicts large amplitude 
primordial gravity waves. We stress that backreaction is not automatically 
detrimental, since corrections often turn beneficial by flattening inflaton potentials and prolonging
inflation instead of shortening it \cite{Dong:2010in,Kaloper:2014zba,McAllister:2014mpa,DAmico:2017cda}. 
The flatter potentials however tend to make the spectral index bluer. This seems to narrow
the remaining theory space for these models. 

In this Letter we argue that such a view is far too pessimistic.
Within the recently expounded framework of rollercoaster cosmology \cite{rollercoaster} 
(see also \cite{Cicoli:2014bja,Braglia:2020eai,Tasinato:2020vdk,Fumagalli:2020nvq,Anguelova:2020nzl,Braglia:2020taf}),
monodromy models remain completely viable, fitting the observations perfectly while 
relaxing the theoretical pressure  from large field
variations. More importantly, they are very predictive, producing primordial gravity 
waves within reach of both the future CMB searches 
such as LiteBIRD and CMB-S4, and
the shorter scale instruments such as LISA. The reason is that the early stage of 
inflation can be completely interrupted, ending with a rapid
reheating and then followed by another stage involving a completely different, second inflaton \cite{rollercoaster}. 

If the first stage lasts some $30$ -- $40$ efolds, the flattened 
monodromy models with $V \sim \phi^p$, $p \lesssim 1/2$
\cite{Dong:2010in,Kaloper:2014zba,McAllister:2014mpa,DAmico:2017cda} easily produce the 
spectra of scalar and tensor perturbations completely within the current observational limits, 
with $n_s \sim 0.965$ and $r \la 0.06$ (depending on $p$ and number of efolds, we can 
have a slightly larger upper bound for $r$). The flattening of the potential reduces $r$.
% on top of the flattening suppression, comes from 
%lower energy density early in inflation since  the early stage lasts shorter than $\sim 60$ efolds. 
Since the predictions are calculated at the pivot point of $N \sim 30$ -- $40$ efolds before the 
end of the {\it first stage} of inflation, $n_s$ is more red. 
Hence both $n_s$ and $r$ move towards the observationally favored regime. 
Note also that the flattening dynamics which generates very shallow potentials activates other irrelevant operators,
including higher derivative corrections to EFT 
\cite{Dong:2010in,Kaloper:2014zba,McAllister:2014mpa,DAmico:2017cda}. These 
terms yield equilateral nongaussianity $f_{NL}^{eq} \simeq {\cal O}(1)$ at CMB scales \cite{DAmico:2017cda}, 
again potentially accessible to detection. This also yields a lower bound on 
$r$, which makes the models quite predictive.  Note further 
that additional stronger nongaussianities can be produced at shorter 
scales, during the interruption between the two stages of inflation,
when the fields turn in the field space \cite{nongauss} (see~\cite{Welling:2018ttr} for 
an extensive review). Those are however not directly observable at present. 

Further, since the inflatons are axions, they generically come from dimensional reduction 
of higher rank forms which have anomalous couplings. 
As a result the UV completion of inflatons yields the usual dimension-5 
operators $\propto \phi F_{\mu\nu}  \tilde F^{\mu\nu}/f$ in four dimensions, 
where $\tilde F_{\mu\nu} = \epsilon_{\mu\nu\lambda\sigma} F^{\lambda\sigma}/2$. When the 
field $\phi$ depends on time, such
as an inflaton in the early stage, and the time dependence is reasonably fast, such as 
near the end of the first stage of inflation,
one of the gauge field helicities becomes tachyonic at large wavelengths \cite{nkaloper} and the 
tachyonic instability leads to a 
rapid, nonperturbative generation of the field $F$. In turn $F$ sources chiral gravity waves \cite{pinky,evazald} 
with an amplitude much greater than the
non-chiral gravity waves generated by the standard metric fluctuations \cite{stary,misao,rollercoaster}. 
The dominant chiral gravity waves become 
superhorizon just before the end of the first stage of inflation, and their enhanced 
amplitude remains frozen during the short epoch of matter domination preceding the second stage of 
inflation, which provides additional $20$ -- $30$ efolds \cite{rollercoaster}. During the last stage, the wavelength
of the frozen chiral tensors stretches to the range of $10^8$ km which make them accessible to LISA. 

Thus it would appear that the issues normally interpreted as theoretical challenges to inflation 
might instead be viewed as aspects of 
naturalness influenced by quantum gravity. Quantum gravity forces a modification 
of the flat space naturalness arguments; 
however the effect of these modifications need not be detrimental. On the contrary, 
monodromy inflatons with restricted field range, 
due to e.g. strong 
coupling effects, nevertheless naturally realize long inflation by using multiple inflatons working in unison. 
It turns out that in this sense these models are not only natural, but also more 
predictive than uninterrupted inflation, yielding detectable signals
over a broad range of wavelengths. 

\section{Whither Double-coaster?}

In the rest of this work we will specialize to a simple two-field model in which inflation is realized 
in two stages, connected by a phase during 
which the original inflaton field oscillates, so that the effective equation of state of the 
universe is approximately $w=0$, like CDM. Such models 
naturally occur in monodromy constructions, involving multiple fields with little hierarchies 
between their masses, and flattened effective 
potentials at large field ranges. The two-field potential we use is
\be 
    V(\phi_1, \phi_2) = M_1^4 \[ \(1 + \frac{\phi_1^2}{\mu_1^2}\)^\frac{p_1}{2} - 1 \]
    + M_2^4 \[ \(1 + \frac{\phi_2^2}{\mu_2^2}\)^\frac{p_2}{2} - 1 \] \, .
\label{eq:Vdouble}
\ee
The scales $\mu_1, \mu_2$ normalizing the fields are both $\calO(0.1 \Mpl)$. We take $M_2 \la M_1$. 

Here we assume both fields $\phi_i$ are axions arising from truncating $p$-form gauge potentials in 
string theory compactifications. This immediately gives the mass scales $\mu_i\sim f_i$  linked to 
the axion decay constants. These in turn are bounded by $f \sim M_{Pl}/(M_s L)^q\lesssim M_{Pl}$ 
where $L$ is the size of the compactification cycle giving rise to the relevant axion, and $M_s$ the 
string scale \cite{Banks:2003sx,Svrcek:2006yi}. Hence, $f_i$  typically fall in the range 
$10^{-2}M_{Pl}\lesssim f\lesssim M_{Pl}$, justifying our choice of $\mu_i$. The scales $M_i$ typically arise 
from warping effects or dilution of energy densities with inverse powers of extra dimension volumes 
(see e.g. section 4.1 in~\cite{Dias:2018koa} for a summary). These are either power-law and/or exponentially 
sensitive to the underlying microscopic parameters of a string compactification, and the axions $\phi_i$ arise 
from two different sectors of a given model. Hence generically $M_1\neq M_2$ and thus without 
loss of generality $M_2 \la M_1$.

\begin{figure}[ht]
    \centering
    \includegraphics[scale=.8]{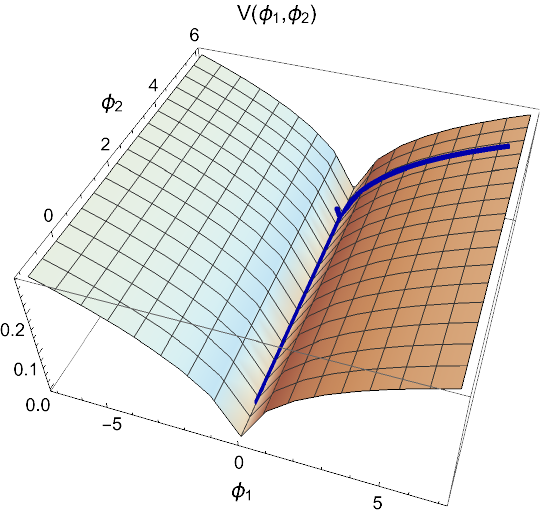}
    \caption{Two-field potential $V(\phi_1, \phi_2)$ for the model in eq.~\eqref{eq:Vdouble}.
    Here $M_1 = 0.5 \Mpl$, $M_2/M_1 = 0.1$, $p_1 = 2/5$, $p_2 = 1$, $\mu_1 = \mu_2 = 0.5$. 
    The blue curve depicts a
    typical two-stage inflationary trajectory, where the field $\phi_1$ 
    slides down the slope first, oscillates while decaying 
    near the bottom of the ``gutter", and then $\phi_2$ starts to move along the ``gutter". }
    \label{fig1}
\end{figure}

These choices decouple 
the early inflationary trajectory from the late inflationary dynamics. Therefore we can study the dynamic 
as a sequence of two consecutive single-field stages. We can view this as a 
very simple realization of Nflation \cite{nflation},
with a larger mass gap between the two inflatons, where one originally 
dominates but falls out of slow roll well before the other field
does. The potential is illustrated in Fig. \ref{fig1}. 

First, we determine the spectrum of perturbations on large scales. 
To do so, we solve the equations of motion for the inflaton and calculate the 
scalar spectral index $n_s$ and the tensor-to-scalar ratio 
$r$ at $N_e$ efolds before the end of the first stage, 
when the scales we see today in the Cosmic Microwave Background (CMB) and 
Large Scale Structure (LSS) leave the horizon. This means, we are using the effective potential during the first stage
\be 
    V_{eff}(\phi_1) = 
    M_1^4 \[ \(1 + \frac{\phi_1^2}{\mu_1^2}\)^\frac{p_1}{2} - 1 \] \simeq M_1^4 \[ \(\frac{\phi_1}{\mu_1}\)^{p_1} - 1 \]  \, ,
\label{eq:Vdoublesin}
\ee
which in the plot of Fig. \ref{fig1} corresponds to fixing $\phi_2$ at 
some value and rolling down the hill towards the ridge 
at a $\phi_2 \simeq {\rm const.}$, thanks to $\phi_2 \gg \mu_2$. 
Explicitly, we define the end of the first stage by $\eps = -\dot{H}/H^2 \simeq 1$, which corresponds to roughly 
$\phi_1(end) \simeq \mu_1$ for our values of $\mu_k$, and use the 
slow-roll expressions $n_s - 1 = 2 \eta_V - 6 \eps_V$, $r = 16 \eps_V$ 
where $\eps_V = \partial_\phi V_{eff}(\phi_1)^2/(2 V_{eff}(\phi_1)^2)$, 
$\eta_V = \partial_\phi^2 V_{eff}(\phi_1)/V_{eff}(\phi_1)$, calculated at a value of $\phi_1$ when 
\be
N_e = \int_{\phi_1}^{\phi_1(end)} d\phi \, \frac{H}{\dot \phi} 
\simeq \(\frac{\phi_1^2}{2p_1 M_{Pl}^2}\)\[1 - \frac{2}{2-p_1} \(\frac{\mu_1}{\phi_1}\)^{p_1}\] \, 
\label{efolds}
\ee%
efolds before the end of the first inflationary stage\footnote{Other terms may correct (\ref{efolds}) 
near the end of the first stage of inflation, such as the inflaton mixing with a dark U(1), to 
appear below. We work in the limit where those corrections can be 
ignored. In particular, the error in neglecting the contribution of the mixing with a 
U(1) to (\ref{efolds}) is a fraction of an efold.}. As $\mu_1$ decreases, $N_e$ increases for fixed $p_1$ and $\phi_1$. 
At a first glance this might look surprising since $V_{eff}$ increases as 
$\mu_1$ decreases. However, as $\mu_1$ is scaled 
down the field ventures to flatter regions, and hence inflation runs longer. 
So the spectral index shifts slightly towards less red values. 
With the hierarchy between inflaton scales, this suppresses nongaussianities  
during the initial stage since the dynamics is dominated by one of 
the inflatons, such that its trajectory is almost straight in the field space 
\cite{nongauss}. We solve the equations numerically. 
\begin{figure}[ht]
    \centering
    \includegraphics[scale=0.8]{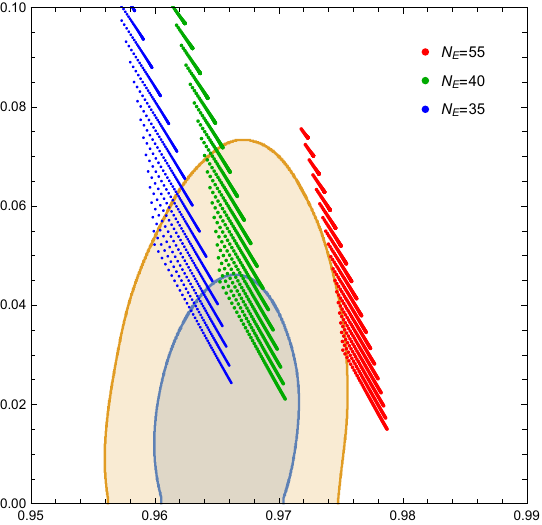}
    \caption{Tensor-to-scalar ratio versus spectral index for double monodromy, compared to 
    the data of~\cite{Ade:2018gkx}.
    We show predictions for the parameter ranges $0.01 \leq \mu_1 \leq 1$, $0.1 \leq p_1 \leq 1$. 
    The lines in the direction of decreasing $r$ and increasing $n_s$ correspond to varying 
    $\mu_1$ at fixed $p_1$; as $\mu_1$ 
    decreases, $N_e$ is larger since the potential is flatter at a fixed $\phi$ and so $n_s$ 
    increases. This also lowers $r$ 
    since a flatter potential implies a smaller $\epsilon$. The lines at approximately 
    constant $n_s$ correspond to varying $p_1$ at fixed $\mu_1$.
    The important point however is that since the efold clock resets at the end 
    of first stage, $n_s$ starts out smaller, and hence is slightly redder
    than in smooth inflation, fitting the data better. This is seen by the comparison of the three 
    cases $N_e = 35, 40, 55$ (where $N_e = 55$ uses a single stage of uninterrupted inflation).}
\label{fig:nsrplot}
\end{figure}

The results are shown in Fig. \ref{fig:nsrplot}.
We display our results similarly to \cite{Dias:2018koa}, plotting predictions for different choices of parameters, 
at three choices of efolds before the end of the first stage of inflation: $N_e = 55, 40, 35$.
It is apparent that double monodromy inflation is fully compatible with the data.
An interesting prediction of this whole class of models is that $r \ga 0.02$, which is well within the reach 
of near future B-modes searches 
(see, e.g. \cite{cmbs4}). 

Restricting $\Delta \phi$'s to be at most a few $M_{Pl}$, as per the current lore about large field
variations (e.g. \cite{ArkaniHamed:2006dz,Ooguri:2006in,Hebecker:2015zss}), we see that 
the individual stages are pretty short. 
For example, for $x = \phi_{max}/M_{Pl} \sim 2$ -- $3$,  and with our values of 
$\mu$, Eq. (\ref{efolds}) yields 
$N_{max} \simeq \(\frac{x^2}{2p}\)\[1 - \frac{1}{1-p/2} \(\frac{1}{2x}\)^{p}\]$. For e.g. $p \sim 0.1$ 
these initial values readily yield $N_{max} \sim 20$ -- $45$, and 
$\Delta \phi$ during inflation is at most $2$ -- $3 \, M_{Pl}$ only.
With slightly larger $p$ the stages of inflation are shorter. Assuming a uniform distribution 
of field values, it is quite generic to obtain a multistage model which, with initial
value of $\phi$'s saturating at $\phi_{max}$ will yield a longer earlier stage and a 
shorter later stage, realizing a double-coaster
of $\sim 40$ and $\sim 20$ stages separated by a brief epoch of matter domination 
supported by the decay of the early inflaton. The price
to pay is to arrange for the right combination of values of $p_k, \mu_k, M_k$, which seems readily attainable. 

Note that in the first stage, the primordial spectrum of tensors remains at $r \la 0.06$ at all scales at 
which the metric fluctuates, by scale invariance. The transition between different stages 
of inflation would not suddenly make
this contribution jump up at shorter scales as noted 
by \cite{stary,misao,rollercoaster}. This opens the door for the nonperturbative generation 
of chiral tensors, using vector tachyon instability \cite{pinky,evazald}. We now turn to this mechanism. 

\section{F\"ur LISA}

Axion inflatons generically couple to $U(1)$ gauge fields via the standard 
dimension-5 operators $\propto \phi_1  F_{\mu \nu} \tilde{F}^{\mu \nu}/2$.
This can be seen in a particularly simple way in flux 
monodromy models \cite{Kaloper:2008fb,Kaloper:2008qs,Kaloper:2011jz}. 
Imagine that the axion sector
arises from a dimensional reduction of, say, a $4$-form field strength in 
$11D$ Sugra, which has Chern-Simons self-couplings. Ignoring the
volume moduli, and imagining a toroidal compactification for simplicity, 
we see that the $11D$ Lagrangian upon dimensional reduction and truncation to zero modes yields 
\ba
-F_{abcd}^2+ \epsilon_{a_1 \ldots a_{11}} A^{a_1 \ldots} 
F^{a_4 \ldots} F^{a_8 \ldots a_{11}} &\ni& - F^2_{\mu\nu\lambda\sigma} 
- (\partial \phi_1)^2 - \mu \phi_1 \epsilon_{\mu\nu\lambda\sigma} F^{\mu\nu\lambda\sigma} \nonumber \\
&& - \sum_k F^2_{\mu\nu \, (k)} - 
\frac{\phi_1}{f_\phi} \sum_{k,l} \epsilon_{\mu\nu\lambda\sigma} F^{\mu\nu}{}_{(k)}  F^{\lambda\sigma}{}_{(l)} \, ,
\ea
where the first line involves the $4$-form-axion sector, and the second the mixing of 
the axion with the $U(1)$ coming from the reduction
of $F_{abcd}$. Here, $A_{abc}$ is the $3$-form potential, $F = dA$. 
The dimensional normalizations come from different scaling dimensions
of various spins, and emerge after the size of internal cycles are accounted for \cite{Witten:1995ex}.  After 
rotating modes in the $U(1)$ isospace 
and canonically normalizing the $4D$ fields, we see that $\phi_1$ will couple to at 
least one $4D$ $U(1)$ vector field. For simplicity, we take 
only one coupling to be nonzero, and model it with the canonically normalized $4D$ dimension-5 operator
\begin{equation}
    \calL_{\rm int} = - \sqrt{-g} \frac{\phi_1}{4 f_\phi} F_{\mu \nu} \tilde{F}^{\mu \nu} \, ,
\label{eq:coupling}
\end{equation}
where $f_\phi$ is sub-Planckian, and generically of the order of GUT scale 
(see e.g.~\cite{Banks:2003sx,Svrcek:2006yi}). A rolling axion triggers the 
tachyonic instability of one circular polarization of the gauge field \cite{nkaloper}, 
whose exponential production both backreacts on the inflaton and 
produces scalar and tensor perturbations \cite{pinky,evazald}.
A very comprehensive analysis of these effects was 
provided recently in \cite{Domcke:2020zez}. The dynamics is governed by \cite{Domcke:2020zez}
\ba
&&\ddot{\phi}_1 + 3 H \dot{\phi}_1 + \partial_{\phi_1} V(\phi_1) -  \frac{1}{f_\phi} \avg{\vec{E} \cdot \vec{B}} = 0 \, , 
\nonumber \\
&& 3 H^2 = \frac{\dot{\phi}_1^2}{2} + V(\phi_1) + \frac{1}{2} \rho_{EB}  \label{veceqs}  \, , \\
&& A''_{\pm}(\tau, \vk) + \[ k^2 \pm 2 \lam \xi k a H \] A_{\pm}(\tau, \vk) = 0 \, , \nonumber 
\ea
where the dot denotes a derivative w.r.t. $t$, and the prime is a derivative w.r.t. to conformal time $\tau$. 
The $U(1)$ `electric' and `magnetic' fields are defined as usual in the Coulomb gauge, 
$\vec{E} = - \frac{1}{a^2} \frac{\rmd \vec{A}}{\rmd \tau}$, $\vec{B} = \frac{1}{a^2} \vec \nabla \times \vec A$, and 
$\rho_{EB} = (\vec E^2 + \vec B^2)/2$ is the standard U(1) energy density. 
We picked the usual circular helicity basis for $\vec A$ in (\ref{veceqs}).  Further $\lam$ is the sign of $\dot{\phi}_1$ 
and $\xi = \frac{1}{2f_\phi} \frac{d\phi_1}{dN_e}$, where $N_e$ is the efold clock reading, for notational convenience.

\begin{figure}[ht]
    \centering
    \includegraphics[scale=0.8]{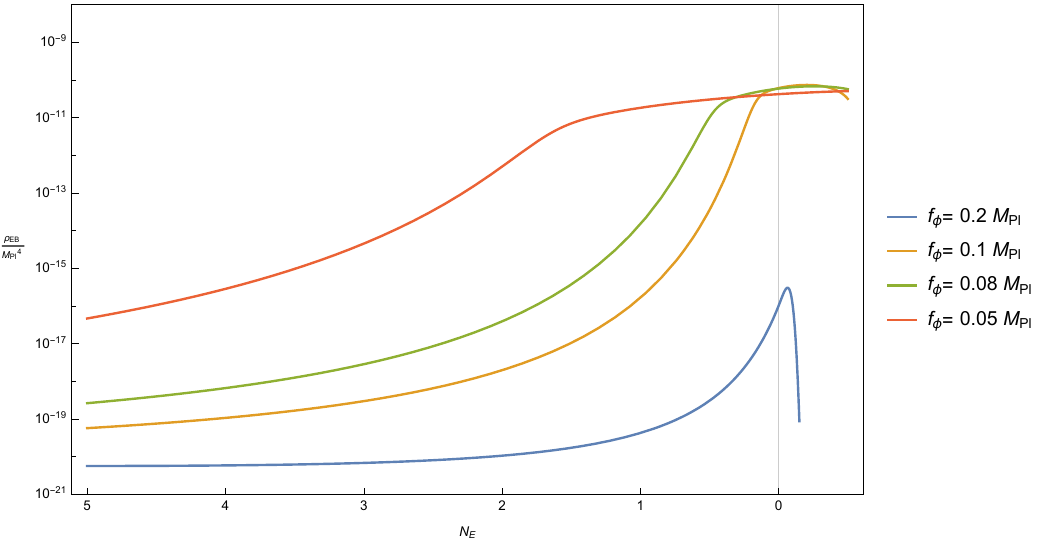}
    \caption{Evolution at the end of the first inflationary stage of the 
    energy density in gauge fields, eq.~\eqref{eq:rhoEB}.
    We denote by $N_e$ the number of efolds before the end of the first 
    stage of inflation, and we normalize the energy density to $\Mpl^4$.
    The parameters of the potential are fixed for convenience to $M_1^4 = 2 \times 10^{-9} \Mpl^4$, $\mu_1 = \Mpl$, 
    $p = 2/5$; at the CMB scales they lead to similar results as $\mu_1 = 0.1 \Mpl$, which fit CMB perfectly. 
    We show results for different couplings $f_{\phi}$ as shown in the legend.
    The vertical line denotes the end of inflation, and the solutions we show are not completely reliable there. 
    Note that the contributions of the vector field are very small until the very end of inflation, 
    when $\phi_1$ moves the fastest. As a result the U(1) production does not affect CMB significantly.}
\label{fig:rhoEB}
\end{figure}

The key ingredient here is the vector field equation. Clearly, if $\dot \phi_1 = 0$, 
it reduces to the standard harmonic oscillator,
where there is no particle production, and the initial population 
dilutes, with physical $\vec E$ and $\vec B$ fields diluting as $1/a^2$. 
On the other hand, when $\dot \phi_1 \ne 0$, and for $2 \xi > k/(a H)$, 
the gauge field helicity $-\lam$ behaves like a tachyon, and is exponentially produced 
by the evolution of $\phi_1$ \cite{Anber:2009ua}. This is the case during inflation. Approximating slow roll with a 
patch of de Sitter where $\xi \simeq {\rm const}.$ yields
\begin{equation}
    A_{-\lam}(\tau, \vk) = \frac{e^{\pi \xi/2}}{\sqrt{2 k}} W_{-i \xi, 1/2}(2 i k \tau) \, , 
    \label{asoln}
\end{equation}
where $W_{k,m}(z)$ is the Whittaker function, and we have imposed the Bunch-Davies vacuum initial conditions.
While this approximation is valid, the average energy density of this field, 
and its Chern-Simons term, can be estimated as, for $\xi \geq 3$,
\begin{equation}
    \rho_{EB} = \avg{\frac{E^2 + B^2}{2}} \simeq 1.3 \cdot 10^{-4} H^4 \frac{e^{2 \pi \xi}}{\xi^3} \, , 
    ~~~~~~~~~   \avg{\vec{E} \cdot \vec{B}} \simeq - 2.4 \cdot 10^{-4} \lam H^4 \frac{e^{2 \pi \xi}}{\xi^4} \, .    
\label{eq:rhoEB}
\end{equation}
We plot the evolution of $ \rho_{EB}$ towards the end of the first inflationary stage, in Fig.~\ref{fig:rhoEB}.

The amplification of the vector field is bounded. First, the instability 
is driven by the slow roll -- the tachyon is an instability of the 
transient background, not of the fundamental theory -- and so the total 
energy deposited in the gauge field cannot exceed the inflaton kinetic energy. 
In particular, as inflation ends and $\phi_1$ settles into its minimum, 
the approximations (\ref{asoln}) and (\ref{eq:rhoEB}) will cease to
apply, with (\ref{asoln}) reducing to the standard harmonic oscillator, as we 
noted above. We will take the transition from one limit to another to
occur close to $\epsilon = - \dot H/H^2 \sim 1$, or in other words, near 
the end of inflation, when $\dot \phi_1$ is almost 
maximal. This sounds counterintuitive, but from there on $\phi_1$ is 
decaying to its minimum, passing through zero
quickly, and completely invalidating the approximations of (\ref{asoln}), (\ref{eq:rhoEB}). 
At this point we will stop the numerical 
integration\footnote{We expect that a more precise description of this transition
can be pursued \`a la WKB, by writing the formal solution to the last of 
Eqs. (\ref{veceqs}) as a contour integral and taking the limits $\xi \gg 1$ and 
$\xi \rightarrow 0$ to match (\ref{asoln}) to the harmonic oscillator amplitude after inflation.}. 
Secondly, while the approximations (\ref{asoln}) and (\ref{eq:rhoEB}) 
are valid, the U(1) vector gauge field energy density will backreact on 
the background evolution and the system will reach some 
equilibrium \cite{Anber:2009ua,Linde:2012bt}. This does not significantly modify early inflation. 
In particular, Fig.~\ref{fig:nsrplot} remains valid as a prediction: early on the metric 
perturbations dominate over the vector-induced gravity waves, since the inflaton is deep in slow roll, 
and inflation is still going on. Essentially, this 
is just decoupling in action. 

Towards the end of the first stage of inflation, 
as the field $\phi_1$ starts to move faster, the larger $\dot \phi_1$ results in a dramatic amplification of $ \rho_{EB}$
cranking up the field strengths by as much as $\sim 10^4 - 10^8$, depending on the value of $f_\phi$.
However the approximations break down at the exit of inflation, 
when $\rho_{EB} \sim \dot \phi_1^2/2$ by energy conservation, and instead the growth in the plots turns 
around, ``plateauing''  near the end of inflation \cite{Domcke:2020zez}.
While we do not zoom in on the specifics of this behavior here, 
the ``plateau" at the end of inflation in Fig. \ref{fig:rhoEB} is interspersed with 
characteristic features since 
as the field $\phi_1$ starts to decay and oscillate, the tachyonic instability of 
the $U(1)$ sector rapidly changes back and forth \cite{DallAgata:2019yrr,Domcke:2020zez} 
\footnote{See also~\cite{Cheng:2015oqa} for earlier related results. 
Again, in Fig. \ref{fig:rhoEB} those effects are smoothed out.}. Those details could 
help tag the signals, warranting a more precise analysis, beyond the numerical `might' we employ here. 

The backreaction of the Chern-Simons term on the inflaton evolution may also alter and amplify scalar perturbations.
In general, \cite{Linde:2012bt,Domcke:2017fix,Garcia-Bellido:2016dkw,Peloso:2016gqs} 
note that this could lead to an enhanced primordial black hole production
after inflation. Yet in the simplest setup which we rely on, with a 
single dark U(1) at scales $\sim 10$ -- $100 \, M_{GUT}$, and without 
ultralight fields charged under U(1), 
it turns out that the enhancement of perturbations due to vector production is limited to about
$\delta \rho/\rho \la {\rm few} \times 10^{-2}$ near the end of the first stage of inflation, and miniscule earlier.  
This will limit distortions and PBH production rates. However, in extended models or with smaller $f_\phi$ those processes could be 
further enhanced.

\begin{figure}[th]
    \centering
    \includegraphics[scale=0.8]{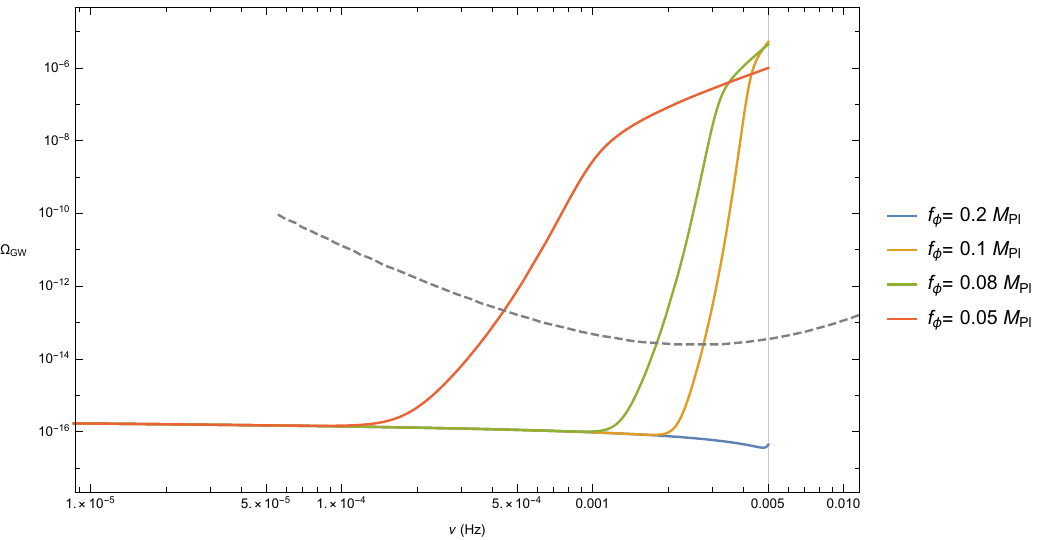}
    \caption{Abundance of gravitational waves as a function of frequency, setting $N_{\rm CMB}=35$.
    The dashed grey line is the sensitivity of LISA, while the horizontal thin line is 
    the bound from the number of relativistic degrees of freedom coming from BBN and CMB. 
    Again, we use $M_1^4 = 2 \times 10^{-9} \Mpl$, $\mu_1 = \Mpl$, $p = 2/5$ for convenience, 
    and as before scan $f_{\phi}$ as shown in the legend. The vertical line again designates the end of inflation,
    beyond which a different approximation is needed.}
\label{fig:OmegaGW}
\end{figure}
Here our main interest are the stochastic gravity waves produced near the end of the first stage of inflation. These  
modes are chiral, and their abundance can be estimated by \cite{Domcke:2017fix}
\begin{equation}
    \Om_{GW} \equiv \frac{\Om_{r,0}}{24} \Delta_T^2
    \simeq \frac{\Om_{r,0}}{12} \(\frac{H}{\pi \Mpl}\)^2
    \( 1 + 4.3 \cdot 10^{-7} \frac{H^2}{\Mpl^2 \xi^6} e^{4 \pi \xi} \) \, ,
    \label{abundance}
\end{equation}
where $\Om_{r,0} = 8.6 \cdot 10^{-5}$ is the radiation abundance today, and the 
terms are evaluated at horizon crossing.
The term in parenthesis adds to the standard tensors produced by metric 
fluctuations in de Sitter space the ``secondary'' production by the 
$U(1)$ vector gauge field. Its contribution tops the metric fluctuations near the end of the first stage of inflation. 

Note, some of the modes produced just before the
end of stage 1 would reenter the horizon during the intermediate matter-dominated stage.
As a result, they would dilute by expansion during that epoch.
Thus the formula for the abundance (\ref{abundance}) should include an extra suppression 
factor. To estimate it, note that for subhorizon modes, the amplitude 
dilutes as $1/\lambda \sim 1/a$, where $a$ is the scale factor. The power in the modes 
is given by the square of the product of the frequency and the amplitude, since these modes behave as harmonic 
oscillators. Hence the suppression factor will be at most $(a_1/a_2)^4$ 
where $a_1$ is the scale factor
at the end of stage 1 of inflation, and $a_2$ the scale factor at the mode re-freezing 
after the beginning of the stage 2 of inflation. This factor is maximized for
the shortest wavelength mode at the end of stage 1, for which $k/a_1 \sim H_1$. 
Since the mode freeze-out gives $k/a_2 
\simeq H_2$, this yields $a_1/a_2 \la H_2/H_1$. For a short intermediate stage  
this would suppress the power by at most a factor of 100 to 1000. 
This suppression would only be a subleading effect, which we ignored here. 
A more precise calculation of these effects would be in order, however, since 
this extra suppression actually helps to evade any conflicts with the BBN bound on gravity waves
depicted in Fig. (\ref{fig:OmegaGW}).

To get an idea how large these modes are, and at which scales they occur,  
we express $\Omega_{GW}$ as a function of 
the frequency observed in instruments at the present time. Since 
the comoving frequency is $\nu = k/(2 \pi)$, we obtain 
the expression for the frequency in terms of the number of efolds before the end of inflation:
\begin{equation}
    N = N_{CMB} + \ln \frac{k_{\rm CMB}}{0.002 \Mpc^{-1}} - 44.9 - \ln \frac{\nu}{10^2 \Hz} \, ,
\end{equation}
where $k_{CMB} = 0.002 \Mpc^{-1}$ is the CMB pivot scale, and $N_{\rm CMB}$ is 
the number of efolds before the end of the first stage of inflation where the CMB scales froze out.
We then re-plot the results of Fig. \ref{fig:rhoEB} in terms of the new independent variable $\nu$.
The results are presented in Fig. \ref{fig:OmegaGW}. Again the approximations which 
we employ are unreliable beyond the end of inflation,
where we terminate the plots. Beyond it the curves would 
bend down. Nevertheless it is clear that these modes are out there for LISA to see. 
Combined with the bounds on $r$ at the CMB scales, which are within reach of 
the future CMB polarization instruments, this makes our 
double-coaster extremely predictive and easy to falsify -- or perhaps, 
confirm. Interestingly, not only would this be a search for a specific 
inflationary model, but also a quest for traces of naturalness on the sky. 

\section{Summary}

To summarize, we have described a very predictive theory 
of axion monodromy inflation. The main difference from the usual realizations of monodromy inflation is 
that inflation is not smooth, but happens in bursts as in rollercoaster cosmology. The example we focused on
involves two axions which feature a little hierarchy between their masses, with initial conditions similar to Nflation.
We motivate the models by combining bounded field ranges with naturalness. 

Due to a little mass hierarchy, however, the axions fall out of slow roll at different times, leading to two stages  
of monodromy inflation separated by
a stage of matter domination, during which the first inflaton oscillates briefly before the second inflaton takes over.
Clearly these masses need to be tuned, although the required tuning does not need to be exceedingly precise. 
If the field ranges are ${\cal O}(M_{Pl})$, and axion potentials are flattened due to being close to the cutoff
and in strong coupling regime, it is straightforward to arrange 
for the two stages of inflation to last ${\cal N} \sim 30 - 40$ and 
${\cal N} \sim 20 - 30$ efolds, respectively. 

This yields the scalar and tensor perturbations at the 
largest scales which fit the CMB perfectly, with $0.02 \la r \la 0.06$, 
in the range of LiteBIRD and CMB-S4 experiments. In addition when the first inflaton 
couples to a hidden sector $U(1)$, which is quite generic in flux 
monodromy models where axions arise from dimensional reduction
of higher rank $p$-forms, there will be an enhanced production of vectors 
near the end of the first stage of inflation. These modes source tensors during the short epoch
of matter domination. These tensors are chiral, with wavelengths at the present time 
in the range of $10^8$ km, and with amplitudes enhanced over the long wavelength modes by vector sources. 
They are a very loud signal for LISA. Hence we find that  
double monodromy inflation easily yields simultaneous 
signals accessible to future gravity wave instruments at different scales. 

\vskip.3cm

{\bf Acknowledgments}: 
We would like to thank A. Lawrence, E. Silverstein, L. Sorbo and especially 
V. Domcke for useful discussions. NK is supported in part by the DOE Grant 
DE-SC0009999. AW is supported by the ERC Consolidator Grant STRINGFLATION 
under the HORIZON 2020 grant agreement no. 647995.

\bibliographystyle{utphys}
\bibliography{references}

\providecommand{\href}[2]{#2}\begingroup\raggedright\begin{thebibliography}{10}

\bibitem{Guth:1980zm}
A.~H. Guth, ``{The Inflationary Universe: A Possible Solution to the Horizon
  and Flatness Problems},''
  \href{http://dx.doi.org/10.1103/PhysRevD.23.347}{{\em Adv. Ser. Astrophys.
  Cosmol.} {\bfseries 3} (1987) 139--148}.

\bibitem{Linde:1981mu}
A.~D. Linde, ``{A New Inflationary Universe Scenario: A Possible Solution of
  the Horizon, Flatness, Homogeneity, Isotropy and Primordial Monopole
  Problems},'' \href{http://dx.doi.org/10.1016/0370-2693(82)91219-9}{{\em Adv.
  Ser. Astrophys. Cosmol.} {\bfseries 3} (1987) 149--153}.

\bibitem{Albrecht:1982wi}
A.~Albrecht and P.~J. Steinhardt, ``{Cosmology for Grand Unified Theories with
  Radiatively Induced Symmetry Breaking},''
  \href{http://dx.doi.org/10.1103/PhysRevLett.48.1220}{{\em Adv. Ser.
  Astrophys. Cosmol.} {\bfseries 3} (1987) 158--161}.

\bibitem{Collins:1972tf}
C.~Collins and S.~Hawking, ``{Why is the Universe isotropic?},''
  \href{http://dx.doi.org/10.1086/151965}{{\em Astrophys. J.} {\bfseries 180}
  (1973) 317--334}.

\bibitem{Kaloper:2011jz}
N.~Kaloper, A.~Lawrence, and L.~Sorbo, ``{An Ignoble Approach to Large Field
  Inflation},'' \href{http://dx.doi.org/10.1088/1475-7516/2011/03/023}{{\em
  JCAP} {\bfseries 1103} (2011) 023},
\href{http://arxiv.org/abs/1101.0026}{{\ttfamily arXiv:1101.0026 [hep-th]}}.
%%CITATION = ARXIV:1101.0026;%%.

\bibitem{Silverstein:2008sg}
E.~Silverstein and A.~Westphal, ``{Monodromy in the CMB: Gravity Waves and
  String Inflation},'' \href{http://dx.doi.org/10.1103/PhysRevD.78.106003}{{\em
  Phys. Rev.} {\bfseries D78} (2008) 106003},
\href{http://arxiv.org/abs/0803.3085}{{\ttfamily arXiv:0803.3085 [hep-th]}}.
%%CITATION = 0803.3085;%%.

\bibitem{McAllister:2008hb}
L.~McAllister, E.~Silverstein, and A.~Westphal, ``{Gravity Waves and Linear
  Inflation from Axion Monodromy},''
  \href{http://dx.doi.org/10.1103/PhysRevD.82.046003}{{\em Phys.Rev.}
  {\bfseries D82} (2010) 046003},
\href{http://arxiv.org/abs/0808.0706}{{\ttfamily arXiv:0808.0706 [hep-th]}}.
%%CITATION = ARXIV:0808.0706;%%.

\bibitem{Kaloper:2008fb}
N.~Kaloper and L.~Sorbo, ``{A Natural Framework for Chaotic Inflation},''
  \href{http://dx.doi.org/10.1103/PhysRevLett.102.121301}{{\em Phys. Rev.
  Lett.} {\bfseries 102} (2009) 121301},
\href{http://arxiv.org/abs/0811.1989}{{\ttfamily arXiv:0811.1989 [hep-th]}}.
%%CITATION = 0811.1989;%%.

\bibitem{Dong:2010in}
X.~Dong, B.~Horn, E.~Silverstein, and A.~Westphal, ``{Simple exercises to
  flatten your potential},''
  \href{http://dx.doi.org/10.1103/PhysRevD.84.026011}{{\em Phys.Rev.}
  {\bfseries D84} (2011) 026011},
\href{http://arxiv.org/abs/1011.4521}{{\ttfamily arXiv:1011.4521 [hep-th]}}.
%%CITATION = ARXIV:1011.4521;%%.

\bibitem{Kaloper:2014zba}
N.~Kaloper and A.~Lawrence, ``{Natural chaotic inflation and ultraviolet
  sensitivity},'' \href{http://dx.doi.org/10.1103/PhysRevD.90.023506}{{\em
  Phys. Rev.} {\bfseries D90} no.~2, (2014) 023506},
\href{http://arxiv.org/abs/1404.2912}{{\ttfamily arXiv:1404.2912 [hep-th]}}.
%%CITATION = ARXIV:1404.2912;%%.

\bibitem{McAllister:2014mpa}
L.~McAllister, E.~Silverstein, A.~Westphal, and T.~Wrase, ``{The Powers of
  Monodromy},'' \href{http://dx.doi.org/10.1007/JHEP09(2014)123}{{\em JHEP}
  {\bfseries 09} (2014) 123},
\href{http://arxiv.org/abs/1405.3652}{{\ttfamily arXiv:1405.3652 [hep-th]}}.
%%CITATION = ARXIV:1405.3652;%%.

\bibitem{DAmico:2017cda}
G.~D'Amico, N.~Kaloper, and A.~Lawrence, ``{Monodromy Inflation in the Strong
  Coupling Regime of the Effective Field Theory},''
  \href{http://dx.doi.org/10.1103/PhysRevLett.121.091301}{{\em Phys. Rev.
  Lett.} {\bfseries 121} no.~9, (2018) 091301},
  \href{http://arxiv.org/abs/1709.07014}{{\ttfamily arXiv:1709.07014
  [hep-th]}}.

\bibitem{rollercoaster}
G.~D'Amico and N.~Kaloper, ``{Rollercoaster Cosmology},''
  \href{http://arxiv.org/abs/2011.09489}{{\ttfamily arXiv:2011.09489
  [hep-th]}}.

\bibitem{Cicoli:2014bja}
M.~Cicoli, S.~Downes, B.~Dutta, F.~G. Pedro, and A.~Westphal, ``{Just enough
  inflation: power spectrum modifications at large scales},''
  \href{http://dx.doi.org/10.1088/1475-7516/2014/12/030}{{\em JCAP} {\bfseries
  12} (2014) 030}, \href{http://arxiv.org/abs/1407.1048}{{\ttfamily
  arXiv:1407.1048 [hep-th]}}.

\bibitem{Braglia:2020eai}
M.~Braglia, D.~K. Hazra, F.~Finelli, G.~F. Smoot, L.~Sriramkumar, and A.~A.
  Starobinsky, ``{Generating PBHs and small-scale GWs in two-field models of
  inflation},'' \href{http://dx.doi.org/10.1088/1475-7516/2020/08/001}{{\em
  JCAP} {\bfseries 08} (2020) 001},
  \href{http://arxiv.org/abs/2005.02895}{{\ttfamily arXiv:2005.02895
  [astro-ph.CO]}}.

\bibitem{Tasinato:2020vdk}
G.~Tasinato, ``{An analytic approach to non-slow-roll inflation},''
  \href{http://arxiv.org/abs/2012.02518}{{\ttfamily arXiv:2012.02518
  [hep-th]}}.

\bibitem{Fumagalli:2020nvq}
J.~Fumagalli, S.~Renaux-Petel, and L.~T. Witkowski, ``{Oscillations in the
  stochastic gravitational wave background from sharp features and particle
  production during inflation},''
  \href{http://arxiv.org/abs/2012.02761}{{\ttfamily arXiv:2012.02761
  [astro-ph.CO]}}.

\bibitem{Anguelova:2020nzl}
L.~Anguelova, ``{On Primordial Black Holes from Rapid Turns in Two-field
  Models},'' \href{http://arxiv.org/abs/2012.03705}{{\ttfamily arXiv:2012.03705
  [hep-th]}}.

\bibitem{Braglia:2020taf}
M.~Braglia, X.~Chen, and D.~K. Hazra, ``{Probing Primordial Features with the
  Stochastic Gravitational Wave Background},''
  \href{http://dx.doi.org/10.1088/1475-7516/2021/03/005}{{\em JCAP} {\bfseries
  03} (2021) 005}, \href{http://arxiv.org/abs/2012.05821}{{\ttfamily
  arXiv:2012.05821 [astro-ph.CO]}}.

\bibitem{nongauss}
C.~Gordon, D.~Wands, B.~A. Bassett, and R.~Maartens, ``{Adiabatic and entropy
  perturbations from inflation},''
  \href{http://dx.doi.org/10.1103/PhysRevD.63.023506}{{\em Phys. Rev. D}
  {\bfseries 63} (2000) 023506},
  \href{http://arxiv.org/abs/astro-ph/0009131}{{\ttfamily
  arXiv:astro-ph/0009131}}.

\bibitem{Welling:2018ttr}
Y.~M. Welling, {\em {Spectroscopy of Two-Field Inflation}}.
\newblock PhD thesis, Leiden U., 11, 2018.

\bibitem{nkaloper}
B.~A. Campbell, N.~Kaloper, R.~Madden, and K.~A. Olive, ``{Physical properties
  of four-dimensional superstring gravity black hole solutions},''
  \href{http://dx.doi.org/10.1016/0550-3213(93)90620-5}{{\em Nucl. Phys. B}
  {\bfseries 399} (1993) 137--168},
  \href{http://arxiv.org/abs/hep-th/9301129}{{\ttfamily arXiv:hep-th/9301129}}.

\bibitem{pinky}
J.~L. Cook and L.~Sorbo, ``{Particle production during inflation and
  gravitational waves detectable by ground-based interferometers},''
  \href{http://dx.doi.org/10.1103/PhysRevD.85.023534}{{\em Phys. Rev. D}
  {\bfseries 85} (2012) 023534},
  \href{http://arxiv.org/abs/1109.0022}{{\ttfamily arXiv:1109.0022
  [astro-ph.CO]}}. [Erratum: Phys.Rev.D 86, 069901 (2012)].

\bibitem{evazald}
L.~Senatore, E.~Silverstein, and M.~Zaldarriaga, ``{New Sources of
  Gravitational Waves during Inflation},''
  \href{http://dx.doi.org/10.1088/1475-7516/2014/08/016}{{\em JCAP} {\bfseries
  08} (2014) 016}, \href{http://arxiv.org/abs/1109.0542}{{\ttfamily
  arXiv:1109.0542 [hep-th]}}.

\bibitem{stary}
D.~Polarski and A.~A. Starobinsky, ``{Spectra of perturbations produced by
  double inflation with an intermediate matter dominated stage},''
  \href{http://dx.doi.org/10.1016/0550-3213(92)90062-G}{{\em Nucl. Phys. B}
  {\bfseries 385} (1992) 623--650}.

\bibitem{misao}
S.~Pi, M.~Sasaki, and Y.-l. Zhang, ``{Primordial Tensor Perturbation in Double
  Inflationary Scenario with a Break},''
  \href{http://dx.doi.org/10.1088/1475-7516/2019/06/049}{{\em JCAP} {\bfseries
  06} (2019) 049}, \href{http://arxiv.org/abs/1904.06304}{{\ttfamily
  arXiv:1904.06304 [gr-qc]}}.

\bibitem{Banks:2003sx}
T.~Banks, M.~Dine, P.~J. Fox, and E.~Gorbatov, ``{On the possibility of large
  axion decay constants},''
  \href{http://dx.doi.org/10.1088/1475-7516/2003/06/001}{{\em JCAP} {\bfseries
  06} (2003) 001}, \href{http://arxiv.org/abs/hep-th/0303252}{{\ttfamily
  arXiv:hep-th/0303252}}.

\bibitem{Svrcek:2006yi}
P.~Svrcek and E.~Witten, ``{Axions In String Theory},''
  \href{http://dx.doi.org/10.1088/1126-6708/2006/06/051}{{\em JHEP} {\bfseries
  06} (2006) 051}, \href{http://arxiv.org/abs/hep-th/0605206}{{\ttfamily
  arXiv:hep-th/0605206}}.

\bibitem{Dias:2018koa}
M.~Dias, J.~Frazer, and A.~Westphal, ``{Inflation as an Information Bottleneck
  - A strategy for identifying universality classes and making robust
  predictions},'' \href{http://dx.doi.org/10.1007/JHEP05(2019)065}{{\em JHEP}
  {\bfseries 05} (2019) 065}, \href{http://arxiv.org/abs/1810.05199}{{\ttfamily
  arXiv:1810.05199 [hep-th]}}.

\bibitem{nflation}
S.~Dimopoulos, S.~Kachru, J.~McGreevy, and J.~G. Wacker, ``{N-flation},''
  \href{http://dx.doi.org/10.1088/1475-7516/2008/08/003}{{\em JCAP} {\bfseries
  08} (2008) 003}, \href{http://arxiv.org/abs/hep-th/0507205}{{\ttfamily
  arXiv:hep-th/0507205}}.

\bibitem{Ade:2018gkx}
{\bfseries BICEP2, Keck Array} Collaboration, P.~Ade {\em et~al.}, ``{BICEP2 /
  Keck Array x: Constraints on Primordial Gravitational Waves using Planck,
  WMAP, and New BICEP2/Keck Observations through the 2015 Season},''
  \href{http://dx.doi.org/10.1103/PhysRevLett.121.221301}{{\em Phys. Rev.
  Lett.} {\bfseries 121} (2018) 221301},
  \href{http://arxiv.org/abs/1810.05216}{{\ttfamily arXiv:1810.05216
  [astro-ph.CO]}}.

\bibitem{cmbs4}
{\bfseries CMB-S4} Collaboration, K.~N. Abazajian {\em et~al.}, ``{CMB-S4
  Science Book, First Edition},''
  \href{http://arxiv.org/abs/1610.02743}{{\ttfamily arXiv:1610.02743
  [astro-ph.CO]}}.

\bibitem{ArkaniHamed:2006dz}
N.~Arkani-Hamed, L.~Motl, A.~Nicolis, and C.~Vafa, ``{The string landscape,
  black holes and gravity as the weakest force},'' {\em JHEP} {\bfseries 06}
  (2007) 060,
\href{http://arxiv.org/abs/hep-th/0601001}{{\ttfamily arXiv:hep-th/0601001}}.
%%CITATION = HEP-TH/0601001;%%.

\bibitem{Ooguri:2006in}
H.~Ooguri and C.~Vafa, ``{On the Geometry of the String Landscape and the
  Swampland},'' \href{http://dx.doi.org/10.1016/j.nuclphysb.2006.10.033}{{\em
  Nucl. Phys.} {\bfseries B766} (2007) 21--33},
\href{http://arxiv.org/abs/hep-th/0605264}{{\ttfamily arXiv:hep-th/0605264
  [hep-th]}}.
%%CITATION = HEP-TH/0605264;%%.

\bibitem{Hebecker:2015zss}
A.~Hebecker, F.~Rompineve, and A.~Westphal, ``{Axion Monodromy and the Weak
  Gravity Conjecture},'' \href{http://dx.doi.org/10.1007/JHEP04(2016)157}{{\em
  JHEP} {\bfseries 04} (2016) 157},
\href{http://arxiv.org/abs/1512.03768}{{\ttfamily arXiv:1512.03768 [hep-th]}}.
%%CITATION = ARXIV:1512.03768;%%.

\bibitem{Kaloper:2008qs}
N.~Kaloper and L.~Sorbo, ``{Where in the String Landscape is Quintessence},''
  \href{http://dx.doi.org/10.1103/PhysRevD.79.043528}{{\em Phys.Rev.}
  {\bfseries D79} (2009) 043528},
\href{http://arxiv.org/abs/0810.5346}{{\ttfamily arXiv:0810.5346 [hep-th]}}.
%%CITATION = ARXIV:0810.5346;%%.

\bibitem{Witten:1995ex}
E.~Witten, ``{String theory dynamics in various dimensions},''
  \href{http://dx.doi.org/10.1016/0550-3213(95)00158-O}{{\em Nucl. Phys. B}
  {\bfseries 443} (1995) 85--126},
  \href{http://arxiv.org/abs/hep-th/9503124}{{\ttfamily arXiv:hep-th/9503124}}.

\bibitem{Domcke:2020zez}
V.~Domcke, V.~Guidetti, Y.~Welling, and A.~Westphal, ``{Resonant backreaction
  in axion inflation},''
  \href{http://dx.doi.org/10.1088/1475-7516/2020/09/009}{{\em JCAP} {\bfseries
  09} (2020) 009}, \href{http://arxiv.org/abs/2002.02952}{{\ttfamily
  arXiv:2002.02952 [astro-ph.CO]}}.

\bibitem{Anber:2009ua}
M.~M. Anber and L.~Sorbo, ``{Naturally inflating on steep potentials through
  electromagnetic dissipation},''
  \href{http://dx.doi.org/10.1103/PhysRevD.81.043534}{{\em Phys.Rev.}
  {\bfseries D81} (2010) 043534},
\href{http://arxiv.org/abs/0908.4089}{{\ttfamily arXiv:0908.4089 [hep-th]}}.
%%CITATION = ARXIV:0908.4089;%%.

\bibitem{Linde:2012bt}
A.~Linde, S.~Mooij, and E.~Pajer, ``{Gauge field production in supergravity
  inflation: Local non-Gaussianity and primordial black holes},''
  \href{http://dx.doi.org/10.1103/PhysRevD.87.103506}{{\em Phys. Rev. D}
  {\bfseries 87} no.~10, (2013) 103506},
  \href{http://arxiv.org/abs/1212.1693}{{\ttfamily arXiv:1212.1693 [hep-th]}}.

\bibitem{DallAgata:2019yrr}
G.~Dall'Agata, S.~Gonz\'alez-Mart\'\i{}n, A.~Papageorgiou, and M.~Peloso,
  ``{Warm dark energy},''
  \href{http://dx.doi.org/10.1088/1475-7516/2020/08/032}{{\em JCAP} {\bfseries
  08} (2020) 032}, \href{http://arxiv.org/abs/1912.09950}{{\ttfamily
  arXiv:1912.09950 [hep-th]}}.

\bibitem{Cheng:2015oqa}
S.-L. Cheng, W.~Lee, and K.-W. Ng, ``{Numerical study of pseudoscalar inflation
  with an axion-gauge field coupling},''
  \href{http://dx.doi.org/10.1103/PhysRevD.93.063510}{{\em Phys. Rev. D}
  {\bfseries 93} no.~6, (2016) 063510},
  \href{http://arxiv.org/abs/1508.00251}{{\ttfamily arXiv:1508.00251
  [astro-ph.CO]}}.

\bibitem{Domcke:2017fix}
V.~Domcke, F.~Muia, M.~Pieroni, and L.~T. Witkowski, ``{PBH dark matter from
  axion inflation},''
  \href{http://dx.doi.org/10.1088/1475-7516/2017/07/048}{{\em JCAP} {\bfseries
  07} (2017) 048}, \href{http://arxiv.org/abs/1704.03464}{{\ttfamily
  arXiv:1704.03464 [astro-ph.CO]}}.

\bibitem{Garcia-Bellido:2016dkw}
J.~Garcia-Bellido, M.~Peloso, and C.~Unal, ``{Gravitational waves at
  interferometer scales and primordial black holes in axion inflation},''
  \href{http://dx.doi.org/10.1088/1475-7516/2016/12/031}{{\em JCAP} {\bfseries
  12} (2016) 031}, \href{http://arxiv.org/abs/1610.03763}{{\ttfamily
  arXiv:1610.03763 [astro-ph.CO]}}.

\bibitem{Peloso:2016gqs}
M.~Peloso, L.~Sorbo, and C.~Unal, ``{Rolling axions during inflation:
  perturbativity and signatures},''
  \href{http://dx.doi.org/10.1088/1475-7516/2016/09/001}{{\em JCAP} {\bfseries
  09} (2016) 001}, \href{http://arxiv.org/abs/1606.00459}{{\ttfamily
  arXiv:1606.00459 [astro-ph.CO]}}.

\end{thebibliography}\endgroup

\end{document}